\documentclass[preprint, aps,prb, amsmath,amssymb, superscriptaddress,showpacs, fleqn,floatfix,footinbib,noeprint]{revtex4-1}
\usepackage[USenglish]{babel}
\usepackage{graphicx}
\usepackage{dcolumn}
\usepackage[colorlinks]{hyperref}
\usepackage[capitalise]{cleveref}

\usepackage[colorinlistoftodos]{todonotes}
\DeclareMathOperator{\sgn}{sgn}
\newcommand{\myud}{\,\mathrm{d}}
\newcommand{\rs}{\ensuremath{r_{\text{s}}}}
\newcommand{\kF}{\ensuremath{k_{\text{F}}}}
\newcommand{\nJ}{\ensuremath{n_{\text{J}}}}
\newcommand{\nAIJ}{\ensuremath{n_{\text{AIJ}}}}
\newcommand{\rmax}{\ensuremath{R}}
\newcommand{\etawithunits}[1]{\ensuremath{\eta_\text{#1}\text{ }(a_0^{-2})}}
\newcommand{\iswitch}{\ensuremath{i_{\text{s}}}}

\begin{document}

\crefname{section}{Sec.}{Sec.}
\crefname{subequations}{Eqs.}{Eqs.}
\Crefname{subequations}{Eqs.}{Eqs.}
\crefformat{subequations}{Eqs.~(#2#1#3)}
\Crefformat{subequations}{Eqs.~(#2#1#3)}
\title{Electronic friction coefficients from the atom-in-jellium model for \texorpdfstring{$Z=1-92$}{Z=1-92}}
\author{Nick Gerrits}
\email{n.gerrits@lic.leidenuniv.nl}
\affiliation{Leiden Institute of Chemistry, Gorlaeus Laboratories, Leiden University, P.O. Box 9502, 2300 RA Leiden, The Netherlands}
\author{J. I\~naki Juaristi}
\affiliation{Departamento de Pol\'imeros y Materiales Avanzados: F\'isica, Qu\'imica y Tecnolog\'ia, Facultad de Qu\'imicas, UPV/EHU, Apartado 1072, 20080 San Sebasti\'an, Spain} 
\affiliation{Centro de F\'isica de Materiales CFM/MPC (CSIC-UPV/EHU), Paseo Manuel de Lardizabal 5, 20018 San Sebasti\'an, Spain}
\affiliation{Donostia International Physics Center (DIPC), Paseo Manuel de Lardizabal 4, 20018 San Sebasti\'an, Spain}
\author{J\"org Meyer}
\email[Corresponding author: ]{j.meyer@chem.leidenuniv.nl}
\affiliation{Leiden Institute of Chemistry, Gorlaeus Laboratories, Leiden University, P.O. Box 9502, 2300 RA Leiden, The Netherlands}
\begin{abstract}
The break-down of the Born-Oppenheimer approximation is an important topic in chemical dynamics on metal surfaces. In this context, the most frequently used work horse is electronic friction theory, commonly relying on friction coefficients obtained from density functional theory calculations from the early 80s based on the atom-in-jellium model. However, results are only available for a limited set of jellium densities and elements ($Z=1-18$). In this paper, these calculations are revisited by investigating the corresponding friction coefficients for the entire periodic table ($Z=1-92$). Furthermore, friction coefficients obtained by including the electron density gradient on the generalized gradient approximation level are presented. Finally, we show that spin polarization and relativistic effects can have sizable effects on these friction coefficients for some elements.
\end{abstract}
\maketitle


\section{Introduction}

Dynamics of surface-molecule reactions are of fundamental importance for a variety of chemical processes, e.g. in heterogeneous catalysis (Haber-Bosch cycle\cite{kozuch2008}).
Fundamentally, the understanding of these dynamics at the atomic scale has so far generally relied on the Born-Oppenheimer (BO) approximation\cite{kroes2008,kroes2012}.
However, in metals, due to the absence of an energy gap for electronic excitations, energy dissipation via electron-hole pair (ehp) excitations could be easily facilitated due to the motion of adsorbate or metal atoms.
Therefore, the validity of the BO-approximation has been questioned for a long time\cite{nieto2006, wodtke2016}.
Even though ehp excitations have been neglected in many theoretical studies in the past, which could also explain experimental data\cite{hellman2006,diaz2009,muzas2012,goikoetxea2014a,luo2016,gerrits2019b,gerrits2019c}, recent studies indicate that ehp excitations can play an important role in the dynamics of molecule-surface reactions\cite{norskov1979,juaristi2008,martin-gondre2012,blanco-rey2014,rittmeyer2015,jiang2016,galparsoro2016,alducin2017,spiering2018,spiering2019}.
For example, vibrational lifetimes of simple diatomic molecules adsorbed on metal surfaces were only explained by going beyond the BO approximation\cite{persson1980,hellsing1984,rantala1986,trail2001,krishna2006,forsblom2007,maurer2016,rittmeyer2015,novko2016,askerka2016,*askerka2017,rittmeyer2017,novko2018,loncaric2019}.
Furthermore, experiments with atomic hydrogen beams have confirmed the importance of ehp excitations\cite{pavanello2013,bunermann2015,janke2015}.

Since the BO-approximation is a very fundamental approximation in theoretical chemistry, going beyond imposes a severe conceptual challenge. 
Alternatively, solving the fully coupled electron-nuclear time-dependent Schr\"{o}dinger equation to completely avoid the BO approximation altogether will remain computationally intractable for the foreseeable future, even for systems with only very few degrees of freedom.
For going beyond the BO approximation, a commonly used approach is combining ab initio molecular dynamics with electronic friction theory \cite{blanco-rey2014,alducin2017}.
Using the local density friction approximation (LDFA) is a way to include the dissipative effect of electron-hole pair excitations in molecular dynamics\cite{juaristi2008}
that is computationally much more convenient than other approaches \cite{maurer2017,spiering2018,spiering2019,zhang2019,zhang2019a,zhang2020}.
Within the LDFA including the independent atom approximation, the so-called electronic friction coefficient $\eta$ is required. 
$\eta$ only depends on the nuclear charge of the moving atom and the electron density of the metal surface at its point-like nucleus or different atoms-in-molecule decompositions of the latter
\cite{rittmeyer2015,novko2015,novko2016a}.
The friction coefficient is obtained by using the atom-in-jellium model, where the atom is embedded in an infinitely extended homogeneous electron gas of that density.
The energy loss in the jellium model is caused by the momentum loss of the nucleus due to the scattering of the electrons from the gas.
The electronic friction coefficient is obtained from the electronic structure of the atom in jellium, which is obtained from density functional theory (DFT)\cite{echenique1981, puska1983}.

The local (spin) density approximation (L(S)DA) for the exchange-correlation functional in DFT is by construction exact for the jellium background, which is why the electronic structure of atoms in jellium has traditionally also been obtained at this level of theory \cite{almbladh1976,zaremba1977,stott1982}.
However, the electronic structure of the atom in jellium is not homogeneous, and thus LDA is not exact.
For jellium spheres containing only a finite number of electrons, quantum Monte Carlo techniques have been employed \cite{duff2007,takada2015,takada2018}.
For infinitely extended jellium on the other hand, going beyond DFT is much more involved \cite{drummond2011} and has never been used for the calculation of electronic friction coefficients.
Earlier work at the generalized gradient approximation (GGA) to DFT has been done in the context of the effective medium theory (EMT) \cite{puska1981,jacobsen1987,jacobsen1996}.
EMT parameters have been obtained from immersion energies calculated with the atom-in-jellium model
\cite{jacobsen1987,puska1991,jacobsen1996}, and both are modified when using GGA instead of LDA \cite{puska1991,nazarov2005}.
However, the effect of employing GGA instead of LDA on the friction coefficients has not been investigated before.
Therefore, friction coefficients are calculated at the GGA level in this work and compared with those obtained with LDA.

Furthermore, spin polarization can affect the value of the friction coefficient for atoms in jellium at low jellium densities, though it is still a matter of discussion whether spin polarization effects should be included within the LDFA scheme when it is applied to molecules\cite{puska1991,luo2016,papanikolaou1993,nazarov2005}.
For carbon it was found that spin-polarized calculations could result in a $70\%$ reduction of the friction coefficient at low jellium density. Nevertheless, this did not alter results for the dissociative chemisorption of methane on Ni(111)\cite{luo2016}. Moreover, a large amount of other elements across the periodic table were found to exhibit a spin moment when spin polarization was allowed within the atom-in-jellium model\cite{nazarov2005,papanikolaou1993}. Furthermore, the jellium can also be spin polarized, to reflect the magnetic moment in ferromagnetic metals\cite{diezmuino2003} and spin friction has also been observed in STM experiments\cite{wolter2012,ouazi2014}. A thorough study into the effect of spin polarization on the friction coefficient will be presented in this paper.

Finally, the atom-in-jellium model is not only important for gas-surface reactions, but also for other kinds of experiments, e.g. analysis of the energy loss of swift (heavy) ions in solids and surfaces\cite{anthony1982, juaristi1999a, juaristi2000, paul2001, alducin2003, winter2003, nazarov2005a, paul2006, nazarov2008, correa2012, roth2017, roth2017a, caro2017, correa2018, ullah2018}. 
However, the tabulated data of Puska and Nieminen\cite{puska1983}, that is commonly used in this context, is limited to the first three, incomplete, rows of the periodic table -- which is insufficient for studies involving energy dissipation of heavier atoms on metal surfaces \cite{liu2006,kisiel2011,rittmeyer2016}.
To extend the amount of elements for which the atom-in-jellium model can be applied, we present here the electronic friction coefficients from hydrogen up to uranium for a variety of jellium densities.
Although it is well known that relativistic effects can influence the electronic structure of heavier free atoms \cite{kotochigova1997,kotochigova1997a}, to the best of our knowledge friction coefficients have not been obtained whilst employing relativistic LDA.
Therefore, we will also investigate the role of relativistic effects for friction coefficients.

The organization of the present paper is as follows: 
In \cref{sec:meth}, first the theory behind the atom-in-jellium model is summarized (\cref{sec:theor:AIJ}) before relativistic extensions (\cref{sec:theor:rel}) and computational details (\cref{sec:comp}) specific to this paper are described. 
In \cref{sec:gga}, a comparison between the results for electronic friction coefficients obtained with LDA and GGA is made.
 \Cref{sec:spinpol} concerns spin polarization. 
Relativistic effects are discussed in \cref{sec:relativistic}. 
Finally, in \cref{sec:conclusion} we summarize the main conclusions of this paper.


\section{Methods}\label{sec:meth}
\subsection{Theory}\label{sec:theor}
Throughout this work Hartree atomic units ($\hbar=e=m_e=1$, $c=\tfrac{1}{\alpha} \approx 137$) are used.

\subsubsection{Non-relativistic atom in jellium}\label{sec:theor:AIJ}

The homogeneous electron gas (jellium) is a model for simple metals that consists of a constant positive background and negative electron charge density resulting in an overall neutral system.
Both densities are characterized by the density parameter $n_0 \geq 0\,a_0^{-3}$ and commonly quantified by the Wigner-Seitz radius $\rs^{-3} = \tfrac{4}{3} \pi n_0$, which is the sphere radius of the mean volume of an electron.

Using spherical coordinates, the radial parts of the corresponding continuum of states are given by spherical Bessel functions $j_l(kr)$.
The (integer) quantum number $l\geq 0$ characterizes the angular momentum, whereas the continuous quantum number $k \in [0;\kF]$ describes the momentum of the state.
The highest occupied state is given by the Fermi energy $E_\text{F}$ and the concomitant Fermi momentum $\kF$:
\begin{equation}
E_\text{F}
\; = \; \frac{1}{2} \kF^2
\; = \; \frac{1}{2} \left( \sqrt[3]{ 3 \pi^2 n_0 } \right)^2
\label{eq:ef}
\end{equation}
Summing over momenta and (an infinite amount of) angular momenta yields the electron probability density of jellium
\begin{equation}
\nJ(r) = 
  \sum\limits_{l} 
    \frac{2l+1}{\pi^2}
      \int_0^{\kF} j_l^2(kr)k^2\myud k \; ,
\label{eq:besselbackground}
\end{equation}
which is constant due to $\sum_l(2l+1)j_l^2(kr)=1$.

Spin-polarized jellium is a simple model for ferromagnetic metals \cite{zong2002}, which introduces homogeneous electron probability densities $\nJ^\sigma(r)$, $\sigma \in \{\uparrow,\downarrow\}$, in the case of collinear spin considered here, such that
\begin{align}
\nJ(r) 
& = \nJ^\uparrow(r) + \nJ^\downarrow(r) \nonumber\\
& = \sum\limits_{\sigma,l} 
	 \frac{2l+1}{2\pi^2}
      \int_0^{\kF^\sigma} j_l^2(kr)k^2\myud k \; .
\label{eq:nup_plus_ndown}
\end{align}
The spin-dependent Fermi momenta are given by
\begin{equation}
\kF^{\uparrow,\downarrow} 
= \sqrt[3]{6 \pi^2 \, \frac{\nJ^{\uparrow,\downarrow}}{1\pm\zeta} }
\quad .
\end{equation}
The strength of the magnetism is characterized by a homogeneous spin polarization
$\zeta = \tfrac{\nJ^\uparrow-\nJ^\downarrow}{n_0}$,
where $\zeta = 0$ corresponds to the original, non-spin-polarized jellium 
($\nJ^\uparrow = \nJ^\downarrow = \tfrac{n_0}{2}$) and $\zeta = 1$ to the ferromagnetic case ($\nJ^\uparrow = n_0$, $\nJ^\downarrow = 0$).
Throughout the rest of this paper, the spin up channel represents the majority spin channel, i.e., $\zeta \geq 0$, $\nJ^\downarrow(r) \leq n_0 \leq \nJ^\uparrow(r)$ and $ \kF^\downarrow \leq \sqrt[3]{ 3 \pi^2 n_0 } \leq \kF^\uparrow $.

In the atom-in-jellium model, homogeniety is destroyed by immersing an atom in a jellium background with density $\rs$. This model can be solved approximately using DFT. Assuming spherical symmetry, the following one-electron Kohn-Sham equations for the radial part of the atom, which is centered at the origin, need to be solved numerically 
\begin{equation}
\left[ -\frac{1}{2r^2} \frac{\partial}{\partial r} \left( r^2 \frac{\partial}{\partial r} \right) + \frac{l(l+1)}{2r^2} + V^\sigma(r) \right] \psi^\sigma(r) 
 = \epsilon^\sigma \, \psi^\sigma(r),
\label{eq:LDA_radial}
\end{equation}
where $\psi^\sigma(r)$ and $\epsilon^\sigma$ are the radial part and the corresponding eigenenergy for the Kohn-Sham orbitals. 
Due to the spherical symmetry these orbitals are $(2l+1)$ degenerate in the (omitted) magnetic quantum number $m$.
The spectrum consists of (localized) bound states $\psi^{\text{b},\sigma}_{n,l}(r)$, 
which are characterized by the main ($n$) and angular ($l$) quantum numbers,
and (delocalized) scattering states $\psi^{\text{sc},\sigma}_{l}(r; k)$, 
which yield the total electron probability density
\begin{align}
\nAIJ(r) & = \, \nAIJ^\uparrow(r) + \nAIJ^\downarrow(r) \nonumber\\[1.5ex]
& = \, \sum_{\sigma,n,l}(2l+1)|\psi^{\text{b},\sigma}_{n,l}(r)|^2 \nonumber\\
& \quad +\sum_{\sigma,l} 
    \frac{2l+1}{2\pi^2} 
    \int_0^{\kF^\sigma} |\psi^{\text{sc},\sigma}_{l}(r; k)|^2 k^2 \myud k 
\; ,
\label{eq:n_DFT}
\end{align}
analogously to \cref{eq:nup_plus_ndown}, where $\nAIJ^\uparrow(r) = \nAIJ^\downarrow(r)$ in the non-spin-polarized case.
The potential $V^\sigma(r)$ in \cref{eq:LDA_radial} is given by
\begin{align}
 V^\sigma(r)=
 & \int\frac{\nAIJ(r')-n_0}{|\mathbf{r}'-\mathbf{r}|}\myud \mathbf{r}' - \frac{Z}{r} \nonumber\\
 & +V^\sigma_{\text{xc}}\left(r;\nAIJ^\uparrow,\nAIJ^\downarrow\right) 
   -V^\sigma_{\text{xc}}\left(r;\nJ^\uparrow,\nJ^\downarrow\right) ,
\label{eq:potential}
\end{align}
where $Z$ is the nuclear charge of the immersed atomic impurity and $V^\sigma_{\text{xc}}$ is the exchange-correlation potential.
Choosing $V^\sigma_{\text{xc}}$ of the jellium background as the zero reference of the potential [as done in \cref{eq:potential}] yields energy eigenvalues $\epsilon^{\text{b},\sigma}_{n,l}<0$ ($0<\epsilon^{\text{sc},\sigma}_{l}(k)<\left(\kF^\sigma\right)^2$) for the bound (scattered) states.
Since $V^\sigma$ depends on the electron distribution, \cref{eq:LDA_radial} needs to be solved self-consistently.

The scattering states are normalized by matching them at the cutoff radius $\rmax$ to their asymptotic limit \cite{duff2007a},
\begin{equation}
 \psi^{\text{sc},\sigma}_{l}(\rmax; k) 
 = \cos\delta^\sigma_l(k) \cdot j_l(k\rmax)
 - \sin\delta^\sigma_l(k) \cdot n_l(k\rmax),
\label{eq:psi_limit}
\end{equation}
where $j_l$ and $n_l$ are the spherical Bessel and Neumann functions, respectively. 
The phase shift $\delta^\sigma_l(k)$ is given by
\begin{equation}
\delta^\sigma_l(k) = 
\tan^{-1}\left(
\frac{ (\ln \psi^{\text{sc},\sigma}_{l})'(\rmax; k) \cdot j_l(k\rmax) - k \cdot j_l'(k\rmax) }
 { (\ln \psi^{\text{sc},\sigma}_{l})'(\rmax; k) \cdot n_l(k\rmax) - k \cdot n_l'(k\rmax) }
 \right),
\label{eq:phase_shift}
\end{equation}
where 
\begin{equation}
(\ln \psi^{\text{sc},\sigma}_{l})'(\rmax; k) 
= \frac{(\psi^{\text{sc},\sigma\;}_{l})'(\rmax; k) }
{\psi^{\text{sc},\sigma}_{l}(\rmax; k)}.
\label{eq:log_deriv_scat}
\end{equation}
The electronic friction coefficient $\eta$ can be calculated from the difference between the phase shifts $\delta_l(\kF)$ of the scattering states at the Fermi energy \cite{ferrell1977,echenique1981}:
\begin{equation}
\eta = \sum_{\sigma,l} 
\frac{(\kF^\sigma)^2}{3 \pi}
(l+1)\sin^2\left(\delta^\sigma_{l+1}(\kF^\sigma)-\delta^\sigma_l(\kF^\sigma)\right).
\label{eq:friction}
\end{equation}
If the jellium background is not spin polarized and the atomic impurity does not induce spin polarization, the summation over the $\sigma$ in \cref{eq:friction} simply yields a factor two, since the phase shifts for spin up and spin down are identical.
Due to the complete screening of the nuclear charge $Z$ by the jellium background, the phase shifts obey the Friedel sum rule\cite{friedel1958},
\begin{equation}
	 \frac{1}{\pi}\sum_{\sigma,l}(2l+1)(\delta^\sigma_l(\kF^\sigma)-\delta^\sigma_l(0))=Z-Z_b,
\label{eq:Friedel}
\end{equation}
with $Z_b$ being the amount of bound electrons.
The atom-induced density of states per unit momentum is given by
\begin{equation}
    \frac{\myud \Delta N^\sigma(k)}{\myud k} = \sum_l \frac{2l+1}{\pi} \frac{\myud \delta^\sigma_l(k)}{\myud k}.
\label{eq:dos}
\end{equation}

\subsubsection{Full and scalar relativistic extension}\label{sec:theor:rel}

\paragraph{RLDA.}
We have extended the atom-in-jellium model to account for relativistic effects.
In the fully-relativistic case the following Kohn-Sham-Dirac radial equations need to be solved \cite{strange1998},
\begin{subequations}
\label{eq:RLDA_radial}
\label[subequations]{eqs:RLDA_radial}
\begin{align}
\frac{\partial g(r)}{\partial r}
 & = -\frac{\kappa+1}{r}g(r) + 2M_\text{R}(r) \, c \, f(r) 
 \label{eq:RLDA_radial_g} \; ,\\
\frac{\partial f(r)}{\partial r}
 & = \frac{V_\text{R}(r) - \epsilon}{c} g(r) + \frac{\kappa - 1}{r} f(r)
 \label{eq:RLDA_radial_f}
\; ,
\end{align}
\end{subequations}
where
\begin{equation}
\label{eq:relat_M}
  M_\text{R}(r) = 1 + \frac{1}{2c^2} \, \left(\epsilon - V_\text{R}(r)\right)
\; .
\end{equation}
The zero of the energy is chosen such that $\epsilon = 0$ describes electrons with zero kinetic energy in the jellium background (i.e., the rest mass of the electron, $c^2$ in present units, has been taken out).
$g(r)$ and $f(r)$ are the radial parts of the large and small components of the two-component Pauli spinors that describe the Kohn-Sham states, respectively.
They are characterized by the relativistic quantum number $\kappa$, that is related to the total angular momentum quantum number $j=l\pm\tfrac{1}{2}$ according to
\begin{equation}
\kappa=
\begin{cases}
 l    & \text{if } j=l-1/2\\
 -l-1 & \text{if } j=l+1/2.
\end{cases}
\label{eq:kappa}
\end{equation}
The potential $V_\text{R}$ in \cref{eq:RLDA_radial,eq:relat_M} has the same form as in \cref{eq:potential}.
In the relativistic local-density approximation (RLDA) used in this paper, a relativistic correction to the (non-relativistic LDA) is included in the exchange-correlation potential \cite{macdonald1979}.
Again, a self-consistent solution is required because $V_\text{R}$ depends on the total electron probability density, which is obtained like in the non-relativistic case [see \cref{eq:n_DFT}] as a sum over bound and scattering states resulting from \cref{eqs:RLDA_radial}.
For the latter, the boundary conditions of the radial parts of the large and small components are \cite{kennedy2004,zabloudil2006}
\begin{subequations}
\label{eq:g_f_limit}
\label[subequations]{eqs:g_f_limit}
\begin{align}
g_{\kappa}^\text{sc}(\rmax; k) =
& \; \cos\delta_\kappa(k) \cdot j_l(k\rmax) 
\nonumber\\
& \;\; - \sin\delta_\kappa(k) \cdot n_l(k\rmax)
\label{eq:g_f_limit_g}\\
\intertext{and}
f_{\kappa}^\text{sc}(\rmax; k) = 
& \; A_{\kappa}(k) \, [\cos\delta_\kappa(k) \cdot j_{\bar{l}}(k\rmax)
\nonumber\\
& \qquad\quad   - \sin\delta_\kappa(k) \cdot n_{\bar{l}}(k\rmax)],
\label{eq:g_f_limit_f}
\end{align}
\end{subequations}
respectively, where $\bar{l} = l -\sgn(\kappa)$ and
\begin{equation}
\label{eq:Akk}
A_{\kappa}(k) = \frac{k c \cdot \sgn({\kappa})}{\epsilon(k)+2c^2}
\; .	
\end{equation}
The phase shift is then\cite{strange1998,kennedy2004,zabloudil2006}
\begin{equation}
\delta_\kappa^{\text{RLDA}}(k) = 
 \tan^{-1} \left(
 \frac{L_{\kappa}(k) \cdot j_l(k \rmax) 
 - A_{\kappa}(k) \cdot j_{\bar{l}}(k \rmax)}
 {L_{\kappa}(k) \cdot n_l(k \rmax) 
 - A_{\kappa}(k) \cdot n_{\bar{l}}(k \rmax)}
 \right),
\label{eq:phaseshift_RLDA}
\end{equation}
where
\begin{equation}
L_{\kappa}(k) 
= \frac{f_{\kappa}^\text{sc}(\rmax; k)}
	   {g_{\kappa}^\text{sc}(\rmax; k)}
\; .
\label{eq:RLDA}
\end{equation}
This yields electronic friction coefficient $\eta_\text{RLDA}$ according to \cref{eq:friction} by summing over $\kappa$ instead of $l$ and $\sigma$.

\paragraph{ScRLDA.}
In addition to the fully relativistic treatment, we have also implemented a scalar-relativistic description according to the approximation proposed by \citet{koelling1977}:
Eliminating the small component and averaging over the spin-orbit components in \cref{eqs:RLDA_radial} leads to
\begin{equation}
\left[ -\frac{1}{2r^2} \frac{\partial}{\partial r} \left(r^2\frac{\partial}{\partial r} \right) 
+ \frac{l(l+1)}{2r^2} + V_\text{ScR}^\sigma(r) \right.
\left. - \frac{1}{4c^2} \frac{\partial V_\text{ScR}^\sigma(r)}{\partial r}\frac{\partial}{\partial r} \right] \frac{\tilde{g}^\sigma(r)}{M_\text{ScR}^\sigma} 
 = \epsilon^\sigma \tilde{g}^\sigma(r)
\label{eq:ScRLDA_radial}
\end{equation}
$M_\text{ScR}$ is defined analogously to \cref{eq:relat_M} using the potential $V_\text{ScR}$.
In the scalar relativistic local-density approximation (ScRLDA), $V_\text{ScR}$ corresponds to $V_\text{R}$, but is based on the electron distribution that is obtained self-consistently with \cref{eq:ScRLDA_radial}. 
The total electron probability density is calculated as before [see \cref{eq:n_DFT}] as a sum over bound and scattering states, which are characterized by the same quantum numbers as in the non-relativistic case.
For $c \rightarrow \infty$ (and thus $M_\text{ScR}^\sigma \rightarrow 1$), \cref{eq:ScRLDA_radial} reduces to the non-relativistic case given by \cref{eq:LDA_radial}.
After substituting the corresponding non-relativistic quantum numbers into \cref{eq:g_f_limit_g}, the boundary conditions for the scattering states are identical to the non-relativistic case given by \cref{eq:psi_limit}.
Consequently, the phase shift is obtained in the same way as in \cref{eq:phase_shift},
\begin{equation}
 \delta^{\text{ScRLDA}, \sigma}_l(k) = 
\tan^{-1}\left( 
 \frac{(\ln \tilde{g}^{\text{sc},\sigma}_{l})'(\rmax; k) \cdot j_l(k\rmax) - k \cdot j_l'(k\rmax)} 
 {(\ln \tilde{g}^{\text{sc},\sigma}_{l})'(\rmax; k) \cdot n_l(k\rmax) - k \cdot n_l'(k\rmax)} 
 \right)
\; ,
\label{eq:phaseshift_ScRLDA}
\end{equation}
where the logarithmic derivative $(\ln \tilde{g}^{\text{sc},\sigma}_{l})'(\rmax;k)$ is defined analogously to \cref{eq:log_deriv_scat}.
The corresponding electronic friction coefficients $\eta_\text{ScRLDA}$ can then be calculated according to \cref{eq:friction} using $\delta^{\text{ScRLDA}, \sigma}_l(\kF^\sigma)$ instead of $\delta^\sigma_{l}(\kF^\sigma)$.

\subsection{Computational Details}\label{sec:comp}

Starting from the atomic solver \texttt{dftatom} by \textcite*{certik2013}, we have developed an in-house code \texttt{LDFAtom} that allows us to numerically solve the atom in jellium model. 
We have coupled our code to \texttt{LibXC}\cite{marques2012}, which implements a large number of commonly-used exchange-correlation functionals.
\texttt{LDFAtom} reproduces the NIST reference for electronic properties of the (free) atoms\cite{kotochigova1997,kotochigova1997a} across the periodic table ($Z=1-92$) using L(S)DA and (Sc)RLDA through \texttt{LibXC} (like \texttt{dftatom} does with its respective direct implementations of these functionals).
We have verified that \texttt{LDFAtom} reproduces immersion energies (see \cref{app:imm_EMT}) for different elements given by \citet{puska1981}, \citet{duff2007} as well as \citet{nazarov2005}.
%
Further numerical details are given in \cref{app:numerics}.

For calculations of friction coefficients at the LDA level, the parametrization by \textcite*{perdew1981} (PZ-LDA) is used, including relativistic corrections suggested by \textcite*{macdonald1979} when needed. 
The GGA according to \textcite*{perdew1996} is used as a representative example for the GGA level.
All the friction coefficients that are discussed in the following section are tabulated in the Supplemental Material\cite{SI}.


\section{\label{sec:results}Results}


\subsection{\label{sec:gga}Generalized gradient approximation}
\begin{figure}
\includegraphics[width=1.0\columnwidth]{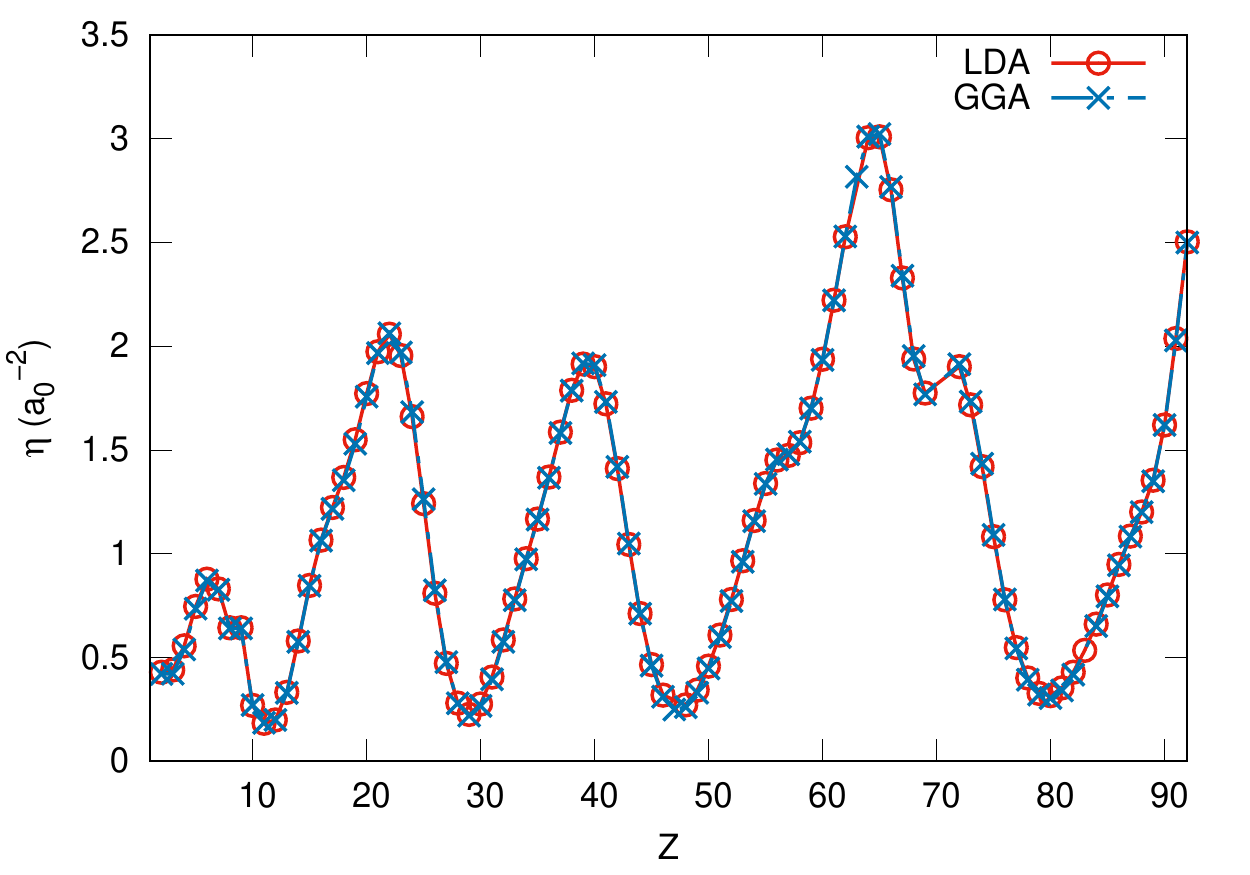}
\caption{\label{fig:qgga}%
The friction coefficients for $Z=1-92$ at $\rs=2$ using LDA (red circles) and GGA (blue crosses). Lines are merely to guide the eye. The numerical data is tabulated in the Supplemental Material\cite{SI}.}
\end{figure}

\Cref{fig:qgga} compares the friction coefficients for $Z=1-92$ at $\rs=2$ using LDA and GGA. We reproduce the results presented by Puska and Nieminen\cite{puska1983} for $Z=1-18$ at various densities and for $Z=1-40$ at $\rs=2$ using LDA. The differences between friction coefficients obtained with LDA and GGA are negligible. This is also observed at other jellium densities. The lack of difference between LDA and GGA is also found for the induced density of states. Since the difference between the induced density of states obtained with LDA and GGA is negligible, it is not surprising that the friction coefficients remain unchanged. 

This is at odds with the fact that previously it has been reported that including the gradient has an influence on the EMT parameters\cite{puska1991}, specifically the neutral sphere radius and cohesive function, within the same atom-in-jellium model. \textcite*{puska1991} have used the GGA parametrization by {Perdew and Wang~\cite{perdew1986,*perdew1989} (PW86)}. Here we confirm to have obtained similar results for the cohesive function using the PBE parametrization.
In general, the neutral sphere radius is larger when using GGA compared to LDA.
Furthermore, the cohesive function is shifted to higher energies (making the cohesive energy larger) and the cohesive function's minimum is at a lower background density compared to LDA.
Since the LDA and GGA yield different immersion energies and potentials, different EMT parameters are obtained\cite{puska1991}.

\begin{figure}
\includegraphics[width=1.0\columnwidth]{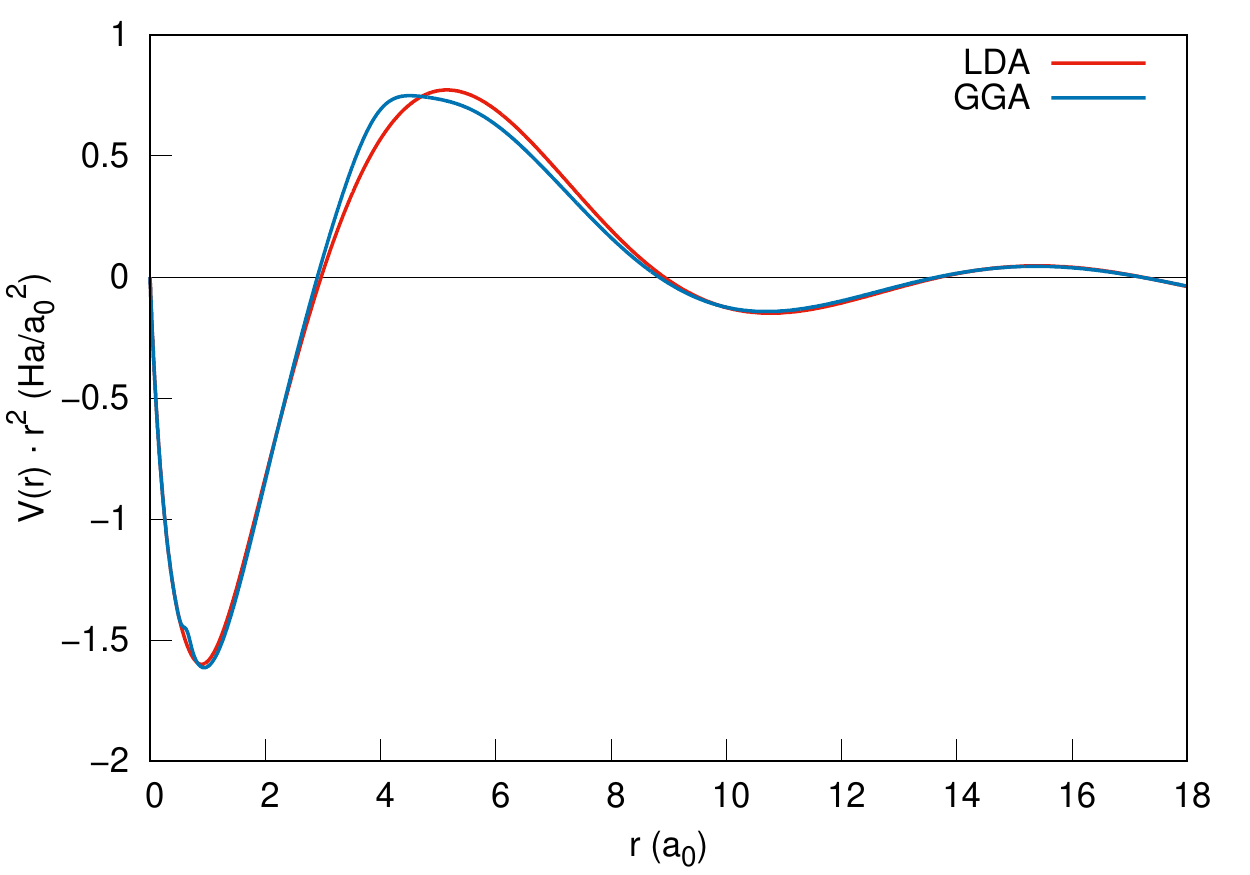}
\caption{\label{fig:ggapotential}%
The total potential [see \cref{eq:potential}] multiplied with $r^2$ for carbon at $\rs=5$ using LDA (red) and GGA (blue).}
\end{figure}

On closer inspection, the main correction of GGA over LDA comes from spatial regions where the reduced density gradient ($\frac{\nabla n(\mathbf{r})}{n(\mathbf{r})^{4/3}}$) is large. This correction is particularly relevant wherever the total electron probability density $n$ is low and its gradient is large -- as is the case in the exponential tail of the free atom electron density at large distances. 
Consequently, the exchange-correlation energy and thus the total energy of the free atom is significantly different.
Since the latter enters the expression of the immersion energy [\cref{eq:immersion}], GGA yields significantly different values for this EMT parameter.

Friction coefficients on the other hand are entirely defined by the potential that enters the Kohn-Sham equations for the atom in jellium [\cref{eq:potential}]. \Cref{fig:ggapotential} compares this potential for LDA and GGA (multiplied with $r^2$) for carbon at $\rs=5$. The differences between the potentials are relatively small and largest in the vicinity of the nucleus. This is not surprising because, unlike for the free atom, the aforementioned decay of the total electron probability density does not occur. Electrons at the jellium's Fermi level hardly notice these differences of the potentials close the nucleus. Consequently, the phase shifts and the concomitant friction coefficient are practically unaffected. Another EMT parameter on the other hand, namely, the neutral sphere radius [\cref{eq:EMT_neutral_sphere_radius}], is very sensitive to changes in the electron probability density close to the nucleus mitigated by the GGA potential and thus significantly affected as shown by \citet{puska1991}.


\subsection{\label{sec:spinpol}Spin polarization}

\begin{figure}
\includegraphics[width=1.0\columnwidth]{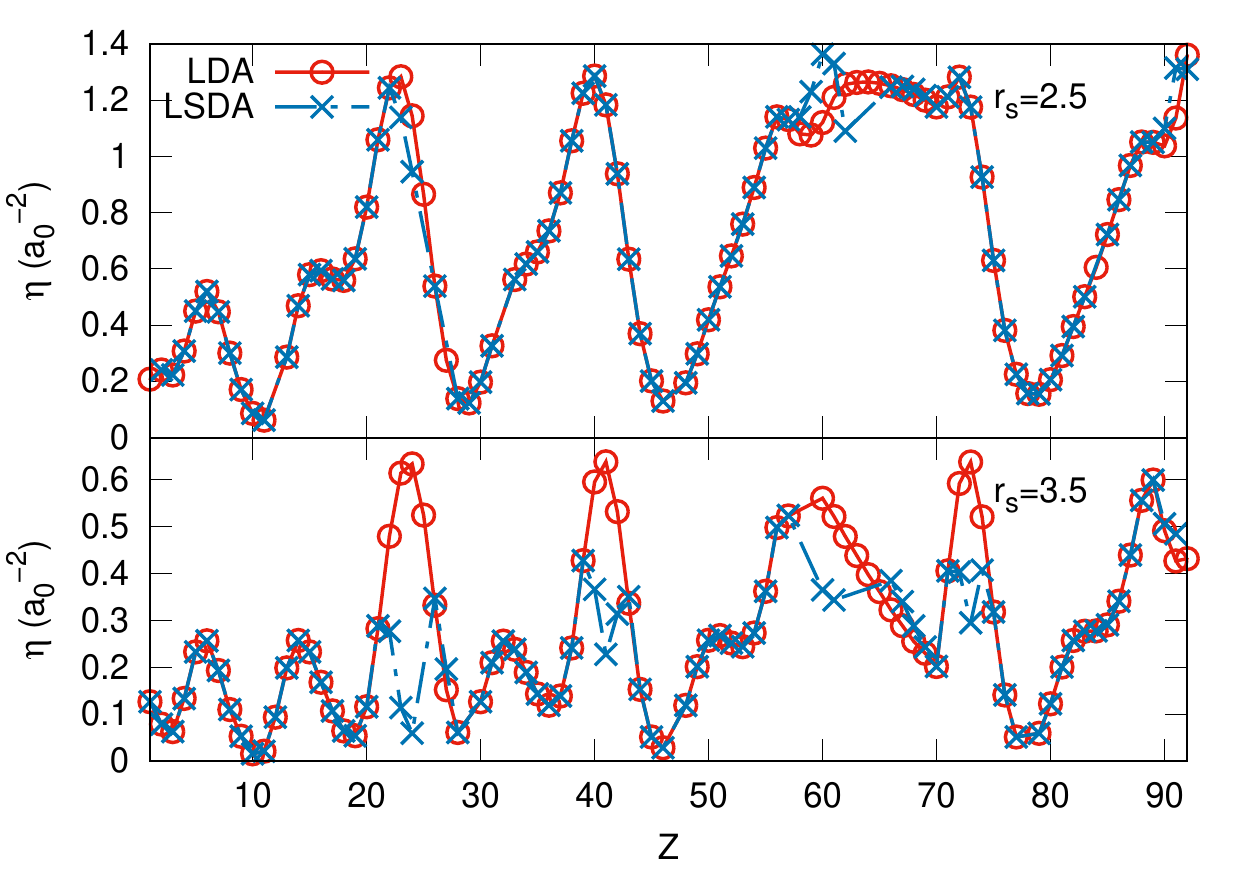}
\caption{\label{fig:qspinrs2535}%
The friction coefficients for $Z=1-92$ at $\rs=2.5$ (top panel) and $\rs=3.5$ (bottom panel). Results obtained with LDA and LSDA are indicated by the red circles and blue crosses, respectively. Lines are merely to guide the eye. The numerical data is tabulated in the Supplemental Material\cite{SI}.}
\end{figure}

\Cref{fig:qspinrs2535} compares the friction coefficients obtained with LDA and LSDA across the periodic table for $\rs=2.5$ and $3.5$. At $\rs=2.5$, spin polarization affects the friction coefficient only for vanadium, chromium, and the majority of the lanthanides and actinides. The differences here are small, ranging from a $15\%$ reduction to $15\%$ increase of the friction coefficients. However, when the background density is lower, spin polarization becomes increasingly more important. Not only are more elements affected by spin polarization, but the differences are relatively larger at lower densities, ranging from a $90\%$ reduction to a $30\%$ increase of the friction coefficients at $\rs=3.5$. Free atoms with a half-filled $d$ or $f$ orbital are the most affected by spin polarization. At even lower densities ($\rs>5$) this effect is also observed for half-filled $p$ orbitals. In general, spin-polarized friction coefficients tend to be lower than non-spin-polarized ones. However, a higher friction coefficient is also possible, seen most prominently for free atoms with an almost empty or completely filled orbital.

\begin{figure}
\includegraphics[width=1.0\columnwidth]{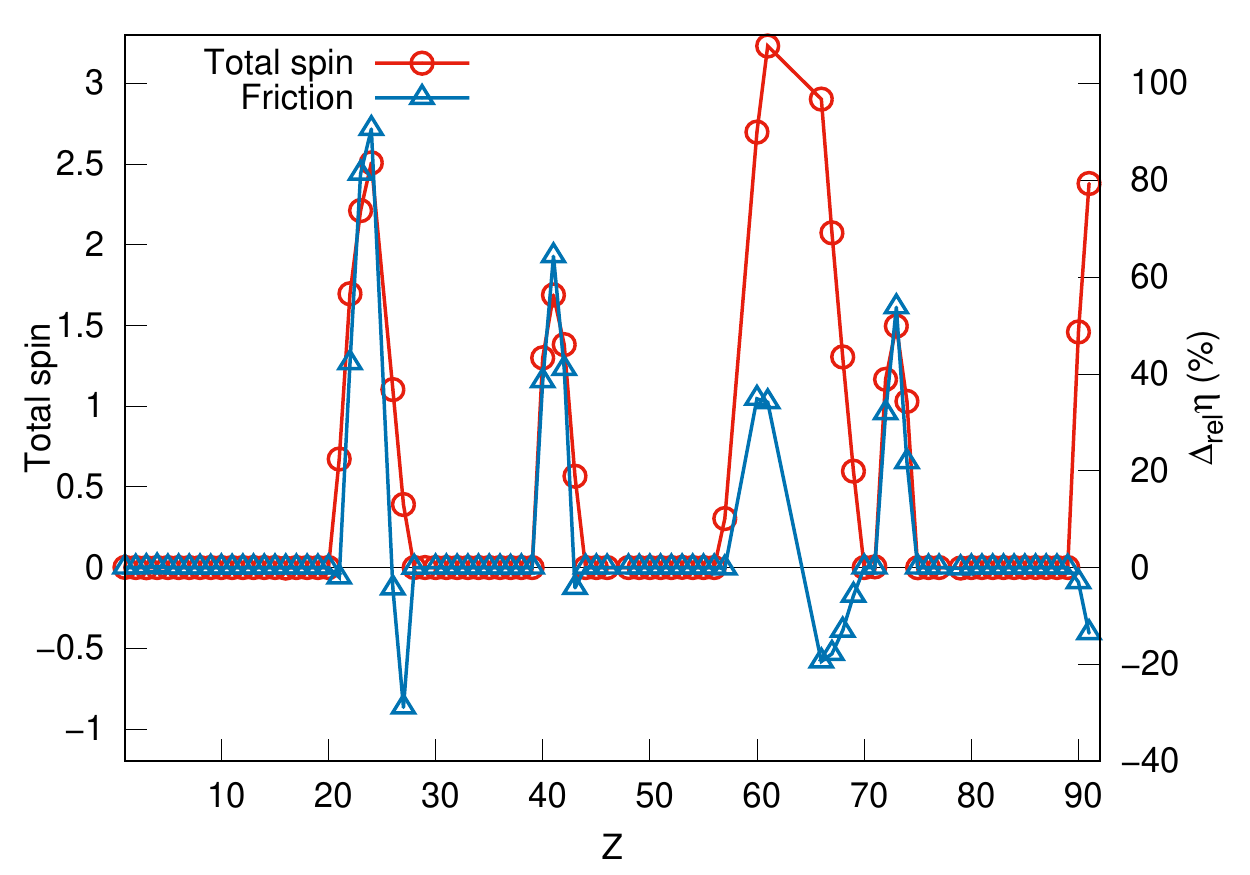}
\caption{\label{fig:spinqdeltars35}
The red circles are the total spin $1/2\int [n^\uparrow(r)-n^\downarrow(r)]\myud r$
for $Z=1-92$ at $\rs=3.5$. 
The blue triangles are the corresponding normalized difference  between the friction coefficients obtained with LDA and LSDA 
$\Delta_\text{rel} \eta = (\eta_{\text{LDA}}-\eta_{\text{LSDA}})/\eta_{\text{LDA}}$.
Lines are merely to guide the eye.}
\end{figure}

To understand what is causing the difference between the friction coefficients, we first compare the trends in total spin and the difference between the friction coefficients due to spin polarization across the periodic table in \cref{fig:spinqdeltars35} at $\rs=3.5$.
The appearance of a (non-zero) total spin coincides with the change in the friction coefficient and is only observed at this density for free atoms with partially filled $d$ and $f$ orbitals.
The total spin is caused by a difference in the amount of scattering spin-up and -down electrons. The maximum total spin found for atoms with a partially filled $f$ orbital is 3.5. Moreover, for atoms with a partially filled $d$ orbital the maximum total spin is 2.5. At lower density ($\rs>5$), a total spin for atoms with a $p$ orbital is also observed, with 1.5 being its maximum value. The maximum total spin that is observed in the scattering states corresponds to half-filled $f$, $d$, and $p$ orbitals, respectively.

\begin{figure}
\includegraphics[width=1.0\columnwidth]{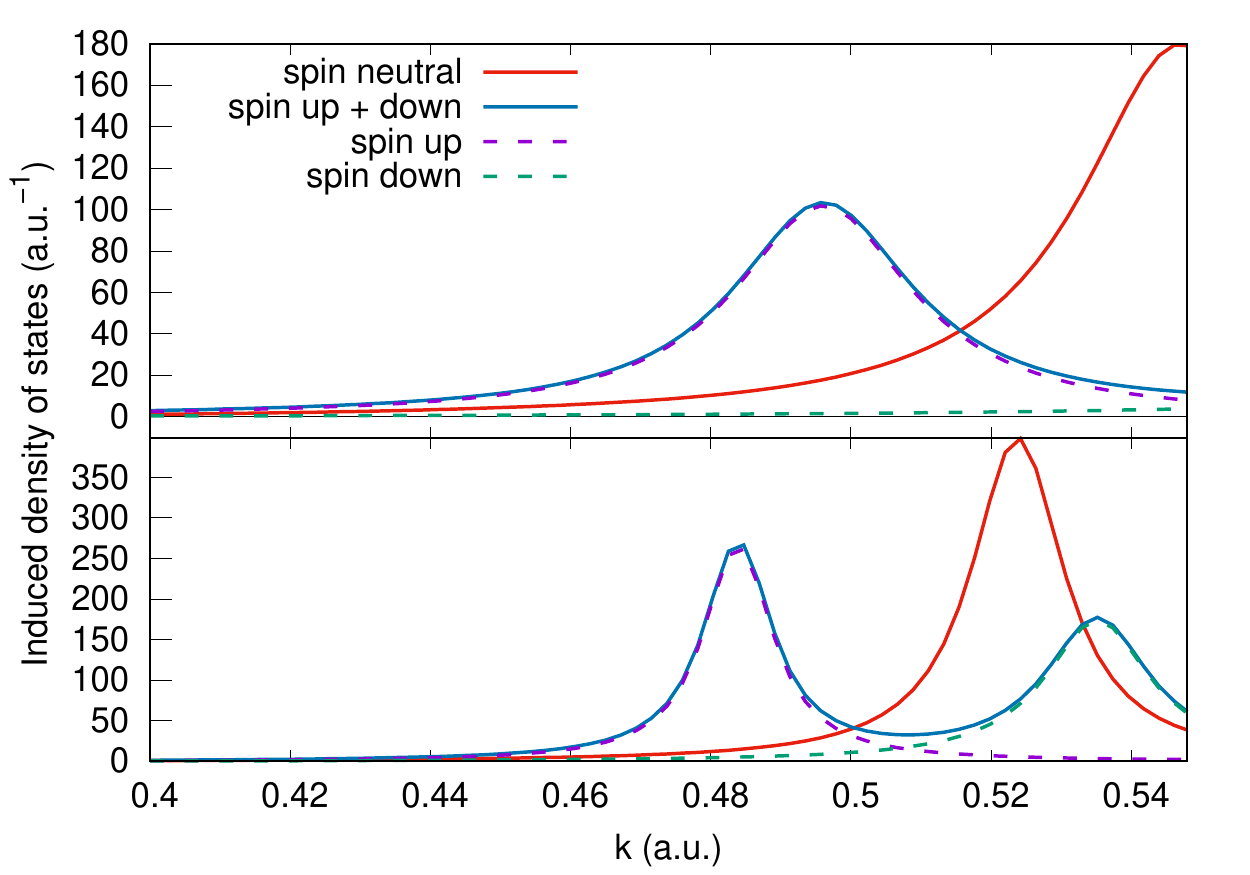}
\caption{\label{fig:dos2327}%
The induced density of states (DOS) of vanadium ($Z=23$, top panel) and cobalt ($Z=27$, bottom panel) at $\rs=3.5$ from $k=0.4$ to $\kF$. DOSs obtained with LDA and LSDA are shown in red and blue, respectively, with the spin up (spin down) channels for LSDA being indicated by dashed purple (green) lines.}
\end{figure}

The appearance of a total spin and its effect on the friction coefficient can be understood by looking at the induced density of states [see \cref{eq:friction,eq:dos}] of vanadium ($Z=23$) and cobalt ($Z=27$) at $\rs=3.5$ in \cref{fig:dos2327} for LDA and LSDA. In these cases, the sharp resonance peak near the Fermi energy corresponds to the $d$ scattering states. A small peak at the bottom of the band is also observed for vanadium, caused by the $s$ scattering states.
The $p$ scattering states do not contribute significantly to the induced density of states. The magnitude of the induced density of states at the Fermi energy relates to the magnitude of the friction coefficient. For example, the reduction of the friction coefficient for vanadium (up to $90\%$) is caused by a split in the sharp resonance peak near the Fermi energy. The spin-up states are lowered in energy, while the spin- down states are higher in energy, causing them to partially be pushed out of the band, effectively lowering the induced density of states at the Fermi energy by $90\%$ and thus a lower friction coefficient is obtained. 

As said before, sometimes spin polarization can also cause an increase in the friction coefficient.
Once more, this can be understood from the induced density of states.
For example, the induced density of states resonance peaks of Cobalt are at a lower energy compared to vanadium.
When the resonance peak of Cobalt is split due to spin polarization, the spin down resonance peak is still within the band since the non-spin-polarized resonance peak for Cobalt is at a significantly lower energy than, e.g., for vanadium.
The spin down resonance peak, being closer to the Fermi energy than the non-spin-polarized resonance peak, causes a higher induced density of states at the Fermi energy (increase of $60\%$) and concomitant larger friction coefficient (increase of $30\%$).
The split in the resonance peak near the Fermi energy is also observed for other elements for which spin polarization yields a total spin.
Which specific scattering states contribute significantly to the induced density of states close to the Fermi energy, varies with elements and densities. Furthermore, using GGA instead of LDA does not produce different results.

\begin{figure}
\includegraphics[width=1.0\columnwidth]{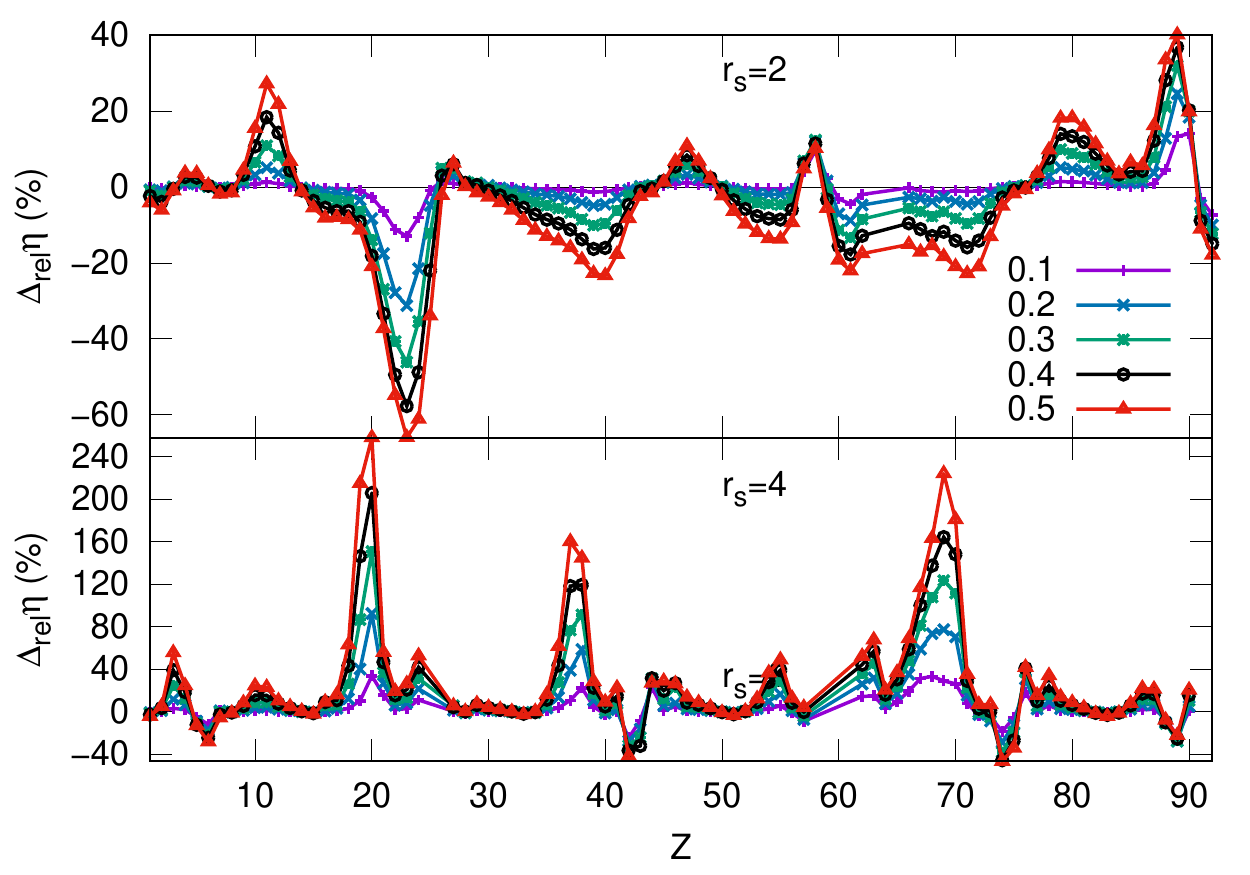}
\caption{\label{fig:qspinpolrs24}%
Difference in friction coefficients between spin-polarized jellium, with $\zeta$ ranging from 0.1 to 0.5, and non-spin-polarized jellium, i.e., $\zeta=0$, obtained with LDA for $Z=1-92$ at $\rs=2$ and $\rs=4$. The lines guide the eye. The numerical data for the friction coefficients is tabulated in the Supplemental Material\cite{SI}.}
\end{figure}

\begin{figure}
\includegraphics[width=1.0\columnwidth]{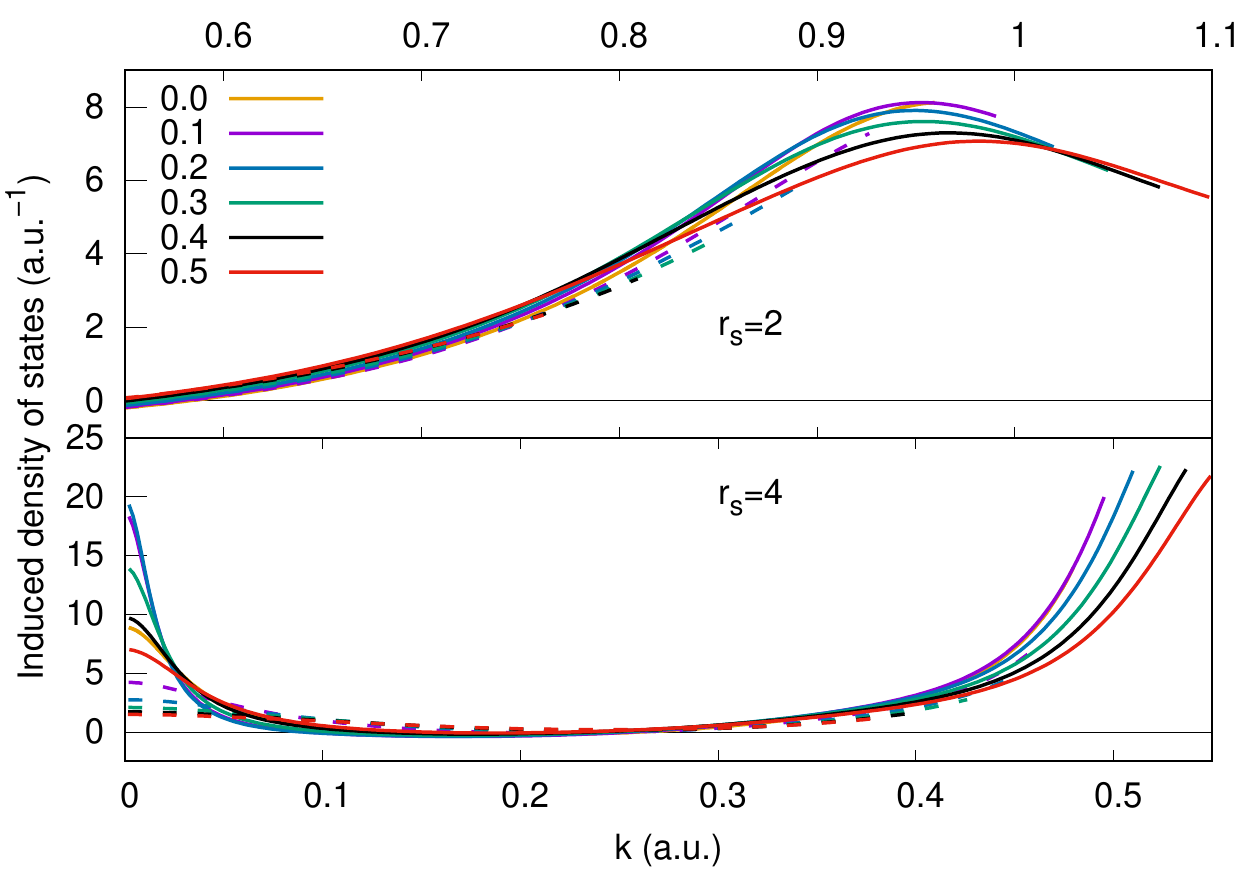}
\caption{\label{fig:dos20rs24}%
The induced density of states obtained with LSDA for $Z=20$ at $\rs=2$ and $\rs=4$ up to $\kF$ for spin-polarized jellium, with $\zeta$ ranging from 0.0 to 0.5. The solid and dashed lines are the spin-up and -down channels, respectively. Note that $\kF$ depends on the jellium density and therefore also on the spin polarization, resulting in different $\kF$ values for the spin-up and -down channels.}
\end{figure}

The differences in the friction coefficients using a spin-polarized jellium compared to a non-spin-polarized jellium ($\zeta=0$) are presented in \cref{fig:qspinpolrs24}. As the spin polarization becomes larger, the differences increase as well.
Whether the friction coefficient increases or reduces is dependent on the element and density, and as such no clear trend is observed.
Again, these differences in the friction coefficients are not caused by the bound states, but by the scattering states.
This can also be seen in \cref{fig:dos20rs24} where the induced density of states for Ca are given at $\rs=2$ and $\rs=4$ for varying spin polarization of the jellium.
Again, we see that the magnitude of the induced density of states at the Fermi energy plays an important role.
In general, if the induced density of states of the spin up channel at the Fermi energy increases, the friction coefficient increases as well, and \textit{vice versa}.
This is similar to what has been observed for \cref{fig:dos2327}.


\subsection{Relativistic effects}\label{sec:relativistic}

\begin{figure}
\includegraphics[width=1.0\columnwidth]{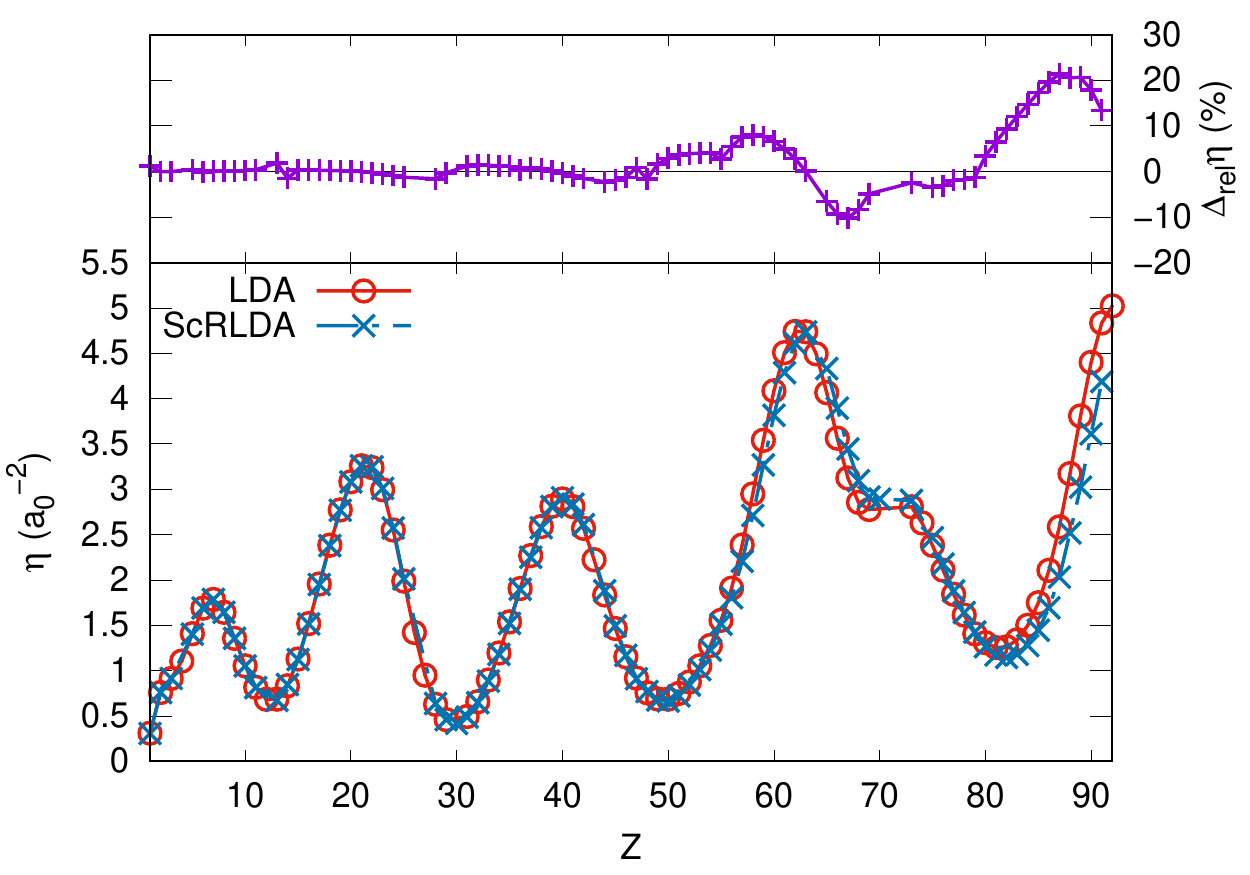}
\caption{\label{fig:qrelativistic}%
The bottom panel shows the friction coefficients $\eta_{\text{LDA}}$ and
$\eta_{\text{ScRLDA}}$ obtained with LDA and ScRLDA, respectively, for $Z=1-92$ at $\rs=1.5$. 
The top panel shows the normalized difference between the friction coefficients using LDA and ScRLDA $\Delta_\text{rel} \eta = (\eta_{\text{LDA}}-\eta_{\text{ScRLDA}})/\eta_{\text{ScRLDA}}$. The red and blue lines are LDA and ScRLDA, respectively. The lines are merely to guide the eye. The numerical data for the friction coefficients is tabulated in the Supplemental Material \cite{SI}.}
\end{figure}

\begin{figure}
\includegraphics[width=1.0\columnwidth]{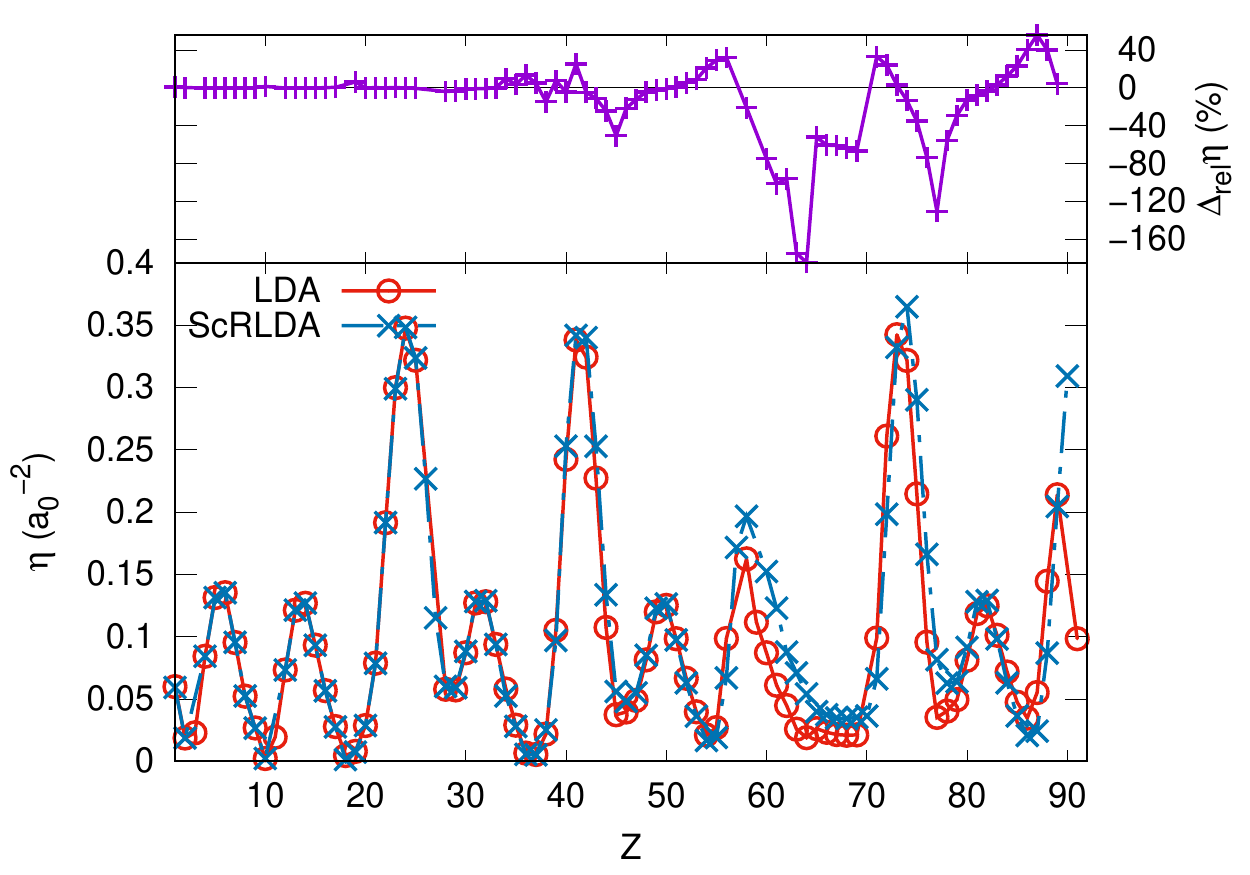}
\caption{\label{fig:qrelativistic5}%
Same as \cref{fig:qrelativistic} at $\rs=5$.}
\end{figure}

The friction coefficients across the periodic table obtained with LDA and ScRLDA are shown in the bottom panel and the corresponding normalized difference in the top panel of \cref{fig:qrelativistic,fig:qrelativistic5} at $\rs=1.5$ and 5, respectively. Relativistic effects influence the friction coefficient significantly for $Z=45$ and heavier atoms, with a maximum difference with respect to the non-relativistic friction coefficients of 20\% at $\rs=1.5$. At lower density, these effects are relatively larger, especially for atoms with partially filled $d$ and $f$ orbitals, ranging from a $40\%$ reduction to $180\%$ increase of the friction coefficient at $\rs=5$. The common trend is that relativistic effects lower the friction coefficient for atoms with partially filled $s$ and $p$ orbitals and increase the friction coefficient for partially filled $d$ and $f$ orbitals.

\begin{figure}
\includegraphics[width=1.0\columnwidth]{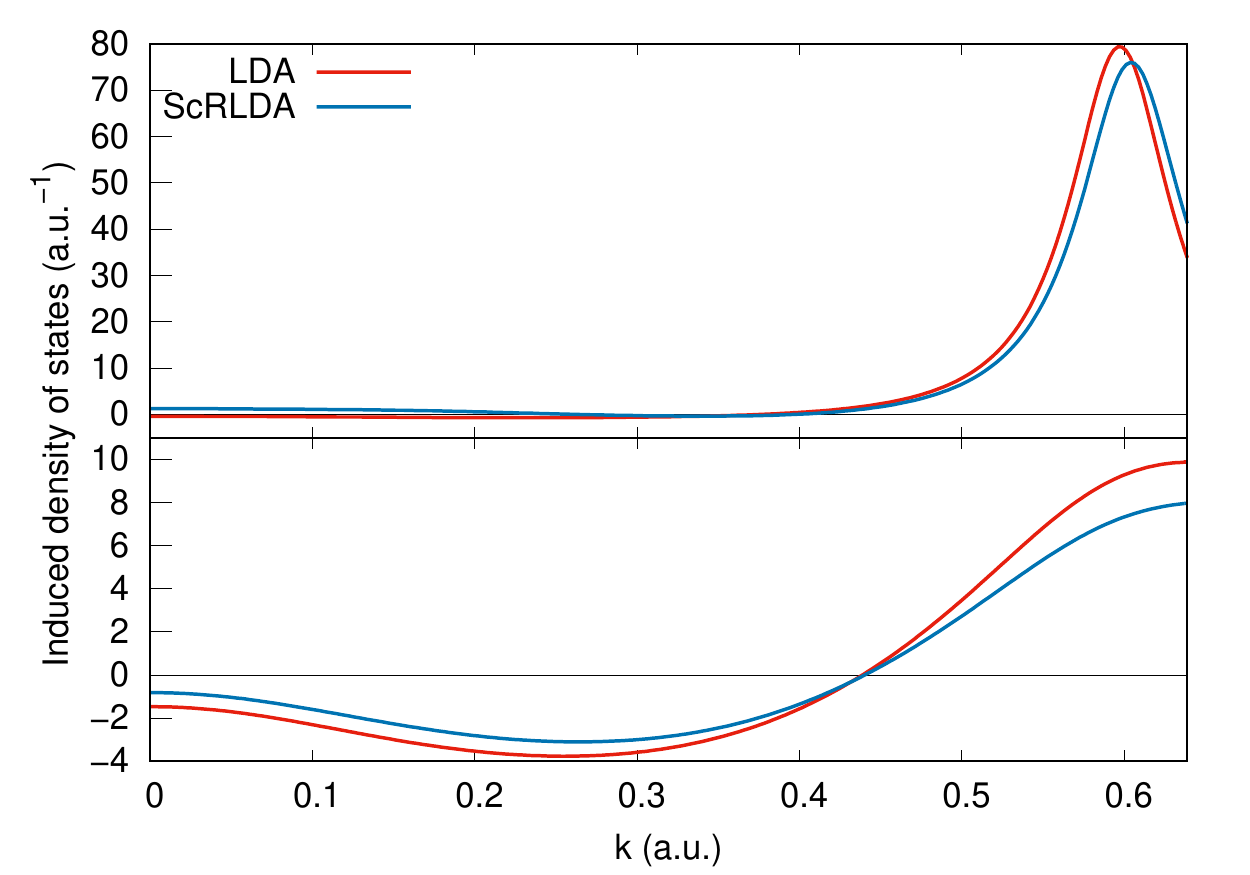}
\caption{\label{fig:dos7486}%
The induced density of states of tungsten ($Z=74$) in the top panel and radon ($Z=86$) in the bottom panel at $\rs=3$ from $k=0$ to $\kF$. The red and blue lines are LDA and ScRLDA.}
\end{figure}

Another look at the induced density of states is required in order to explain the differences caused by relativistic effects. \Cref{fig:dos7486} shows the induced density of states for tungsten ($Z=74$), for which relativistic effects increase the friction coefficient and radon ($Z=86$), which is affected in the opposite way. In general, the induced density of states at low energies is higher due to relativistic effects. Furthermore, the resonance peak near the Fermi energy is lower and is shifted to a higher energy compared to LDA. How this affects the friction coefficient depends on the induced density of states at the Fermi energy. Typically, the induced density of states will be lower within the ScRLDA if the peak is relatively close to the Fermi energy due to the smaller resonance peak, resulting in a reduced friction coefficient. Otherwise, when the resonance peak is at a comparatively lower energy, the shift of the resonance peak increases the induced density of states at the Fermi energy and the friction coefficient. 

Finally, we have a few short remarks on relativistic effects. First, the $6s$ electrons are bound less strongly for $5d$ elements when using ScRLDA compared to LDA. This causes the $6s$ bound states to more easily disappear into the continuum. Nevertheless, this has no significant effect on the friction coefficient. Moreover, spin polarization with the ScRLDA gives the same differences for the friction coefficient as obtained with the LDA. The exception is the $5d$ elements, for which the total spin is partially due to the presence of more bound spin up than spin down electrons originating from the $6s$ orbital, but this results only in a slight increase of the friction coefficients ($<10\%$). This effect was not observed with the LDA. Additionally, in \cref{tab:scrldarlda} friction coefficients are given for a few heavy elements obtained with ScRLDA and RLDA at $\rs=1.5$ and $5$. Fully relativistic calculations did not alter results significantly compared to ScRLDA. At high density ($\rs=1.5$), the differences were smaller than $5\%$. Spin-orbit coupling has a slightly bigger effect ($<10\%$) at low densities ($\rs=5$), but the absolute differences at low densities are small, especially compared to the differences between LDA and ScRLDA.

\begin{table}
\setlength{\tabcolsep}{4pt}
\caption{Friction coefficients for a few heavy elements obtained with ScRLDA and RLDA at $\rs=1.5$ and $5$.}
\label{tab:scrldarlda}
\begin{ruledtabular}
\begin{tabular}{ll | dd | dd}
  \multicolumn{2}{c|}{} & \multicolumn{2}{c|}{$\rs=1.5$} & \multicolumn{2}{c}{$\rs=5$} \\
  \multicolumn{1}{l}{Element} & \multicolumn{1}{c|}{$Z$} &
  \multicolumn{1}{r}{\etawithunits{ScRLDA}} & \multicolumn{1}{r|}{\etawithunits{RLDA}} &  
  \multicolumn{1}{r}{\etawithunits{ScRLDA}} & \multicolumn{1}{r}{\etawithunits{RLDA}} \\
\hline
Pm & 61     & 4.285  & 4.111  & 0.123  & 0.111\\
Dy & 66     & 3.896  & 3.920  & 0.037  & 0.043\\
Re & 75     & 2.466  & 2.447  & 0.290  & 0.278\\
Tl & 81     & 1.172  & 1.152  & 0.128  & 0.124\\
Pb & 82     & 1.144  & 1.120  & 0.129  & 0.127\\
Ra & 88     & 2.518  & 2.400  & 0.087  & 0.076\\
\end{tabular}
\end{ruledtabular}
\end{table}


\section{Conclusions}\label{sec:conclusion}

In this paper, the electronic friction coefficients are calculated using DFT within the atom in jellium model for the entire periodic table ($Z=1-92$) in the range of $\rs=1.5-5$. Furthermore, the influence of a variety of modifications to the widely used atom-in-jellium model on the electronic friction coefficient has been investigated. Using GGA for the xc-functional only affects EMT parameters, the friction coefficient is unaffected. Furthermore, spin polarization can play a significant role, especially for atoms with a half filled $d$ or $f$ orbital. This effect becomes increasingly more dominant when the embedding density is lower and is caused by the polarization of the scattering states. Moreover, having a spin-polarized jellium can heavily influence the friction coefficient, but no clear trend with the atomic number or background density was observed. Finally, at high jellium densities, relativistic effects have only a minor influence on the friction coefficient for heavy elements. However, at low densities these effects are more important, with lanthanides, actinides, and $5d$ elements being affected the most.


\begin{acknowledgments}
N.G. is grateful for his research stay in San Sebastian that has been co-funded by the Erasmus+ programme of the European Union.
J.M. acknowledges financial support from the Netherlands Organisation for Scientific Research (NWO) under VIDI Grant No. 723.014.009.
J.I.J. acknowledges financial support by the Gobierno Vasco-UPV/EHU Project No. IT1246-19, and the Spanish Ministerio de Ciencia e Innovaci{\' o}n [Grant No. PID2019-107396GB-I00/AEI/10.13039/501100011033].
\end{acknowledgments}


\appendix
\section{Immersion energy and effective medium theory parameters}
\label{app:imm_EMT}

The immersion energy, which describes the energy cost or gain of placing an atom in jellium, is obtained from the atom-in-jellium model by taking the energy difference between the atom in jellium, and the pure jellium and free atom \cite{puska1981,duff2007a},
\begin{equation}
    E_{\text{imm}} = E_{\text{AIJ}} - E_{\text{J}} - E_{\text{atom}},
\label{eq:immersion}
\end{equation}
where the energy difference between the atom in jellium and pure jellium can be obtained from a single calculation of the atom in jellium:
\begin{equation}
 E_{\text{AIJ}} - E_{\text{J}} =\Delta T + \Delta E_{\text{coul}} + \Delta E_{\text{xc}} 
\end{equation}
The difference in kinetic energy is
\begin{align}
 \Delta T = 
&   \sum_{\sigma,i}E^\sigma _i - \sum_\sigma 4\pi \int \nAIJ^\sigma(r)V^\sigma(r) r^2 \myud r \nonumber\\
& + \sum_{\sigma,l} \frac{2l+1}{\pi} (\kF^\sigma)^2 \delta_l^\sigma(\kF^\sigma) \nonumber\\
& - \sum_{\sigma,l} \frac{2l+1}{\pi} \int_0^{\kF^\sigma} k\delta_l^\sigma(k)\myud k
\quad.
\end{align}
The difference in Coulomb energy is given by
\begin{align}
\Delta E_{\text{coul}} 
= \int
& \left(\frac{1}{2}\int\frac{\nAIJ(r')-n_0}{|\mathbf{r}-\mathbf{r}'|}
\myud \mathbf{r}' - \frac{Z}{r}\right) \nonumber\\
& \cdot \left( \nAIJ(r)-n_0) \right) \, \myud \mathbf{r}
\end{align}
and the exchange-correlation energy difference is
\begin{align}
\Delta E_{\text{xc}} = \;
&  E_{\text{xc}}\left[ \nAIJ^\uparrow,\nAIJ^\downarrow \right] \nonumber\\
& - E_{\text{xc}}\left[ \nJ^\uparrow,\nJ^\downarrow \right]
  \; + \;  \Delta E_{\text{xc}}^{\text{corr}}
\, ,
\end{align}
where the last term is a correction that accounts for the influence of Friedel oscillations beyond the cut-off radius $\rmax$ \cite{puska1981} -- which are most pronounced for the contribution of $\Delta E_{\text{xc}}$ to $E_{\text{imm}}$. For the verification of our implementation \texttt{LDFAtom} as described in \cref{sec:comp}, we have used the correction originally suggested by \citet{puska1981} 
\begin{align}
\Delta E_{\text{xc}}^{\text{corr}} = \;
& \biggl(
  \varepsilon_{\text{xc}}\left[n^\uparrow,n^\downarrow \right] \bigg|_{n^\uparrow=n^\downarrow=\tfrac{n_0}{2}} \nonumber\\
& \quad + n_0 \frac{\myud \varepsilon_{\text{xc}} \left[n^\uparrow, n^\downarrow=\tfrac{n_0}{2}\right]}
                {\myud n^\uparrow}\bigg|_{n^\uparrow=\tfrac{n_0}{2}}
  \biggr) \nonumber\\
& \cdot \left( Z- 4\pi \int_0^{\rmax}(\nAIJ(r)-n_0) \, r^2 \myud r \right),
\end{align}
to calculate immersion energy curves for various first and second row atoms without spin polarization.

Important parameters for the EMT can be obtained from the atom in jellium model \cite{jacobsen1996}.
The so-called cohesive function,
\begin{align}
E_{\text{c}} = \;
& E_{\text{imm}}(n_0) \nonumber\\
& + n_0 \int_0^s 
\left(
  \int \frac{\nAIJ(r')-n_0}{|\mathbf{r}-\mathbf{r}'|}\myud \mathbf{r}'
  -\frac{Z}{r}
\right) 
\myud \mathbf{r}
\quad,
\end{align}
deserves particular attention in this context.
Here the Coulomb interactions are subtracted from the immersion energy inside the neutral sphere defined by the radius $s$.
The latter is an EMT parameter that is obtained from the electron distribution according to the charge neutrality condition \cite{puska1991}:
\begin{equation}
\label{eq:EMT_neutral_sphere_radius}
4 \pi \int_0^s \nAIJ(r) r^2 \myud r = Z
\quad .
\end{equation}
Minimizing $E_{\text{c}}$ with respect to $n_0$ yields the cohesive energy $E_\text{coh} = | E_{\text{c}}(n_0^\text{coh}) |$ and the concomitant density parameter $n_0^\text{coh}$, which are two very important EMT parameters.
We have implemented the calculation of $E_{\text{c}}$ and $s$ into \texttt{LDFAtom}, but have not made use of it in the scope of this paper.

\section{Numerical details}
\label{app:numerics}

\subsection{Radial Kohn-Sham equations}
\label{app:radial}

In our implementation \texttt{LDFAtom}, the radial Kohn-Sham equations [\cref{eq:LDA_radial,eq:RLDA_radial,eq:ScRLDA_radial}] are solved by rewriting them for the non-relativistic (Schr{\"o}dinger), RLDA and ScRLDA in the form of two coupled first-order differential equations that are completely equivalent to the respective formulation in \cref{sec:meth}. 

Using the substitutions $P(r)=r \, \psi(r)$ and $Q(r)=\psi(r)+ r \, \tfrac{\partial \psi(r)}{\partial r}$ together with \cref{eq:LDA_radial}, the two equations that are solved in the non-relativistic case  are
\begin{subequations}
\begin{align}
\frac{\partial P(r)}{\partial r} & = Q(r) \; , \\
\frac{\partial Q(r)}{\partial r} 
& = 2\left[ \frac{l(l+1)}{2r^2} + V(r) - \epsilon \right]P(r) \; .
\end{align}
\end{subequations}
For the fully relativistic case, the large and small components are substituted by $P(r)=r \, g(r)$ and $Q(r)=r \, f(r)$, respectively, in \cref{eqs:RLDA_radial}, which gives
\begin{subequations}
\begin{align}
\frac{\partial P(r)}{\partial r} 
& = -\frac{\kappa}{r} \, P(r) + \left[ \frac{\epsilon - V_\text{R}(r)}{c} + 2c \right] Q(r) \; , \\
\frac{\partial Q(r)}{\partial r} 
& = -\left[ \frac{\epsilon - V_\text{R}(r)}{c} \right] P(r) + \frac{\kappa}{r} Q(r) \; .
\end{align}
\end{subequations}
Finally, as shown by \citet{koelling1977}, \cref{eq:ScRLDA_radial} in the scalar-relativistic case can be conveniently solved by the substitutions $P = r \, \tilde{g}(r)$ and $Q(r) = \tfrac{r}{2M_\text{ScR}(r)} \tfrac{\partial \tilde{g}(r)}{\partial r}$, resulting in 
\begin{subequations}
\begin{align}
\frac{\partial P(r)}{\partial r} & = 2 \, M_\text{ScR}(r) \, Q(r) + \frac{P(r)}{r} \; , \\
\frac{\partial Q(r)}{\partial r} 
& = -\frac{Q(r)}{r} + \left[ \frac{l(l+1)}{2\,M_\text{ScR}(r)\,r^2} + V_\text{ScR}(r) - \epsilon \right] P(r) \; .
\end{align}
\end{subequations}

\subsection{Grids}
\label{app:grids}

Using the fourth-order Adams-Bashforth integration method \cite{chiou1999} already implemented in \texttt{dftatom}\cite{certik2013}, the equations presented in the preceding \cref{app:radial} are solved on a real-space grid,
\begin{equation}
r_i = r_0 + \frac{r_N-r_0}{\Gamma_{\alpha,\iswitch}(N)} \; ,
\left[ 
\Gamma_{\alpha,\iswitch}(i) - \frac{N-i}{N} \Gamma_{\alpha,\iswitch}(0)
\right]
,
\end{equation}
with $i \in \{0,1,2, \ldots, N\}$ and where
\begin{equation}
\Gamma_{\alpha,x_0}(x) = - \ln(G_{-\alpha,x_0}(x))
\end{equation}
is based on the logistic function:
\begin{equation}
G_{\alpha,x_0}(x) = \frac{1}{1+\exp{(-\alpha(x-x_0))}}
,
\end{equation}
This grid enables adequate sampling near the atomic impurity at the origin because the grid points being logarithmically distributed for $r_0 \leq r_i < r_{\iswitch}$.
For $r_{\iswitch} < r_i \leq r_N$, grid points become more and more equidistant, which adequately samples the long-range part at large distances from the impurity
where perturbation of the jellium has (almost) decayed.
We have found empirically by extensive convergence tests that $\alpha=36 a_0^{-1}$, $\iswitch = \lfloor \tfrac{2}{5} N\rceil$, and $N=6000$ provide a very accurate solution of all calculated properties.
After introducing analytic continuations of the spherical Bessel functions $j_l(kr)$ for small arguments ($ kr \leq 10^{-7}$), we have set $r_0 = 10^{-7} a_0$.
$r_N = \rmax$ has been varied individually for each atom in a range from $18 a_0$ to $28 a_0$ until the Friedel sum rule \cref{eq:Friedel} is numerically fulfilled within $10^{-4}$ in each case.

A sufficient number of angular momenta ($l_\text{max}$) needs to be included in the calculation of the scattering states, which is ensured by mandating 
$\left|\tfrac{\nJ(r_0)}{\nJ(r_N)} - 1 \right| < 10^{-6}$
in a separate calculation for the unperturbed jellium background [see \cref{eq:besselbackground}]. Integrations over $k$ [like e.g. in \cref{eq:n_DFT}] are performed with an equidistant grid of 250 points.

\subsection{Self-consistent solution}

For the initial guess of the atom in jellium, the self-consistent density of the free atom is added to the background density of the jellium.
The mixing between self-consistent field (SCF) cycles is performed with a limited memory version of Broyden's second method\cite{broyden1973,johnson1988b,vanderotten2005}.
The self-consistency is evaluated by checking the convergence of the Kohn-Sham effective potential
the concomitant eigenenergies.
For the former, the Euclidian norm of each spin component of the potential [see \cref{eq:potential}]
\begin{equation}
\lVert V_\text{begin}^\sigma(r) \rVert_2 = 
  \sqrt{4 \pi \int_0^R \left( V_\text{begin}^\sigma(r) \right)^2 r^2 \myud r}
\end{equation}
is calculated at the beginning of each SCF cycle.
Likewise, after the potential has been updated to $V_\text{end}(r)$ at the end of each SCF cycle, the Euclidean norm of the difference with respect to $V_\text{begin}(r)$ is calculated.
If 
$\tfrac{ \lVert V_\text{end}^\sigma(r) - V_\text{begin}^\sigma(r) \rVert_2 }
 { \lVert V_\text{begin}^\sigma(r) \rVert_2 } < 10^{-6}$
for both spin channels, then the potential is considered to be (sufficiently) self-consistent.
For the Kohn-Sham eigenenergies, only the largest difference between the current and previous SCF cycle is considered and only when the potential already fulfills the aforementioned self-consistency criterion.
When this difference is smaller than $5*10^{-6} \, a_0\cdot\text{Ha}$, the eigenenergies are considered to be self-consistent as well and convergence is achieved, i.e., the ground-state solution is obtained.

Weakly bound states can cause calculations not to reach self-consistency. 
This is caused by the appearance and subsequent disappearance of bound states into the continuum between SCF cycles due to the close proximity of these states to the bottom of the continuum at 0~$\text{Ha}$ (for the energy zero chosen in \texttt{LDFAtom}, see \cref{sec:theor:AIJ}).
To stabilize the SCF convergence -- i.e., for purely numerical convenience and without any physical meaning -- a broadening scheme is introduced for the occupation of such weakly bound states using a Fermi-Dirac distribution:
\begin{equation}
f^\text{FD}(\epsilon_{nl}) 
= \frac{2l+1}{\exp(\epsilon_{nl}/\epsilon_\text{B})+1}
  \quad .
\end{equation}
Here $f^\text{FD}(\epsilon_{nl})$ is the occupation number of the bound Kohn-Sham state with energy $\epsilon^\sigma_{n,l}$ and $\epsilon_\text{B}$ is the broadening parameter. The bound state search is stopped when $\epsilon^\sigma_{n,l}>5\epsilon_\text{B}$.
We have used $10^{-3}\, \text{Ha}< \epsilon_\text{B} < 10^{-2}\,\text{Ha}$ 
and confirmed that this does not affect the friction coefficients significantly. 
However, even with this approach, SCF convergence could not be achieved in some cases, mainly $d$ and $f$ elements at low jellium densities.

\bibliography{main}

\end{document}